\documentclass[useAMS,usenatbib,twocolumn]{mnras}

\usepackage{natbib}
\usepackage{graphicx}
\usepackage{bm}
\usepackage{amssymb}
\usepackage{amsmath}
\usepackage{verbatim}
\usepackage{mathrsfs}
\usepackage{amsfonts}
\usepackage{latexsym}
\usepackage{color}
\usepackage{units}
\usepackage{array}
\usepackage{relsize}
\usepackage{geometry}
\usepackage{hyperref}
\usepackage{array,booktabs,arydshln,xcolor}
\usepackage{multirow}
\usepackage[utf8]{inputenc}

\usepackage{epsfig}
\usepackage{graphicx,subfigure}
\usepackage{slashbox}
\usepackage{booktabs}

\begin{document}
           
\title[Verlinde's emergent gravity, MOND and dSph's]{Verlinde's emergent gravity versus MOND and the case of dwarf spheroidals}

\author[Alberto Diez-Tejedor et al]{Alberto Diez-Tejedor$^{1}$\thanks{E-mail: alberto.diez@fisica.ugto.mx}, Alma X. Gonzalez-Morales$^{1,2}$, Gustavo Niz$^{1}$\\
 $^{1}$Departamento de F\'isica, Divisi\'on de Ciencias e Ingenier\'ias,
Campus Le\'on, Universidad de Guanajuato, 37150, Le\'on, M\'exico\\
 $^{2}$Consejo Nacional de Ciencia y Tecnolog\'ia, Av. Insurgentes Sur 1582,
 Colonia Cr\'edito Constructor, Del. Benito Juárez, 03940,\\ M\'exico D.F. M\'exico}
 \date{\today}
\maketitle

\begin{abstract}
In a recent paper, Erik Verlinde has developed the interesting possibility that spacetime and gravity may emerge from the entangled structure of an underlying 
microscopic theory. In this picture, dark matter arises as a response to the standard model of particle physics from the delocalized degrees of freedom that 
build up the dark energy component of the Universe. Dark matter physics is then regulated by a characteristic acceleration scale $a_0$, identified with the radius of the 
(quasi)-de Sitter universe we inhabit. For a point particle matter source, or outside an extended spherically symmetric object, MOND's empirical fitting formula is recovered. However, Verlinde's theory 
critically departs from MOND when considering the inner structure of galaxies, differing by a factor of 2 at the centre of a regular massive body. For illustration, we use 
the eight classical dwarf spheroidal satellites of the Milky Way. These objects are perfect testbeds for the model given their  approximate spherical symmetry, measured 
kinematics, and identified missing mass. We show that, without the assumption of a maximal deformation, Verlinde's theory can fit the velocity dispersion profile in dwarf 
spheroidals with no further need of an extra dark particle component. If a maximal deformation is considered, the theory leads to mass-to-light ratios that
are marginally larger than expected from stellar population and formation history studies. We also compare our results with the recent 
phenomenological interpolating MOND function of McGaugh {\it et al}, and find a departure that, for these galaxies, is consistent with the scatter in current observations.

\end{abstract}


\section{Introduction and summary of results}\label{sec.introduction}

The dark matter (DM) problem remains one of the big puzzles in modern cosmology~\citep{Bertone:2004pz, DM}. It is commonly accepted that this component consists of a new 
particle, not included in the standard model sector, which is expected to interact so weakly with the visible matter that, at present, it has not been detected in 
accelerators nor through direct or indirect astrophysical probes [see however~\cite{2009arXiv0912.3828V}, \cite{Bernabei:2015msa}, and~\cite{Parsons:2015isa} for positive 
claimed signals]. It is noteworthy that all solid evidence pointing to the existence of this mysterious component comes from observations involving the 
gravitational interaction in some way or another. Therefore, one may also consider that the mass discrepancy suggesting the existence of a DM particle could be a 
consequence of a poor understanding of gravity at the relevant scales.

\begin{figure}
\includegraphics[width=\columnwidth]{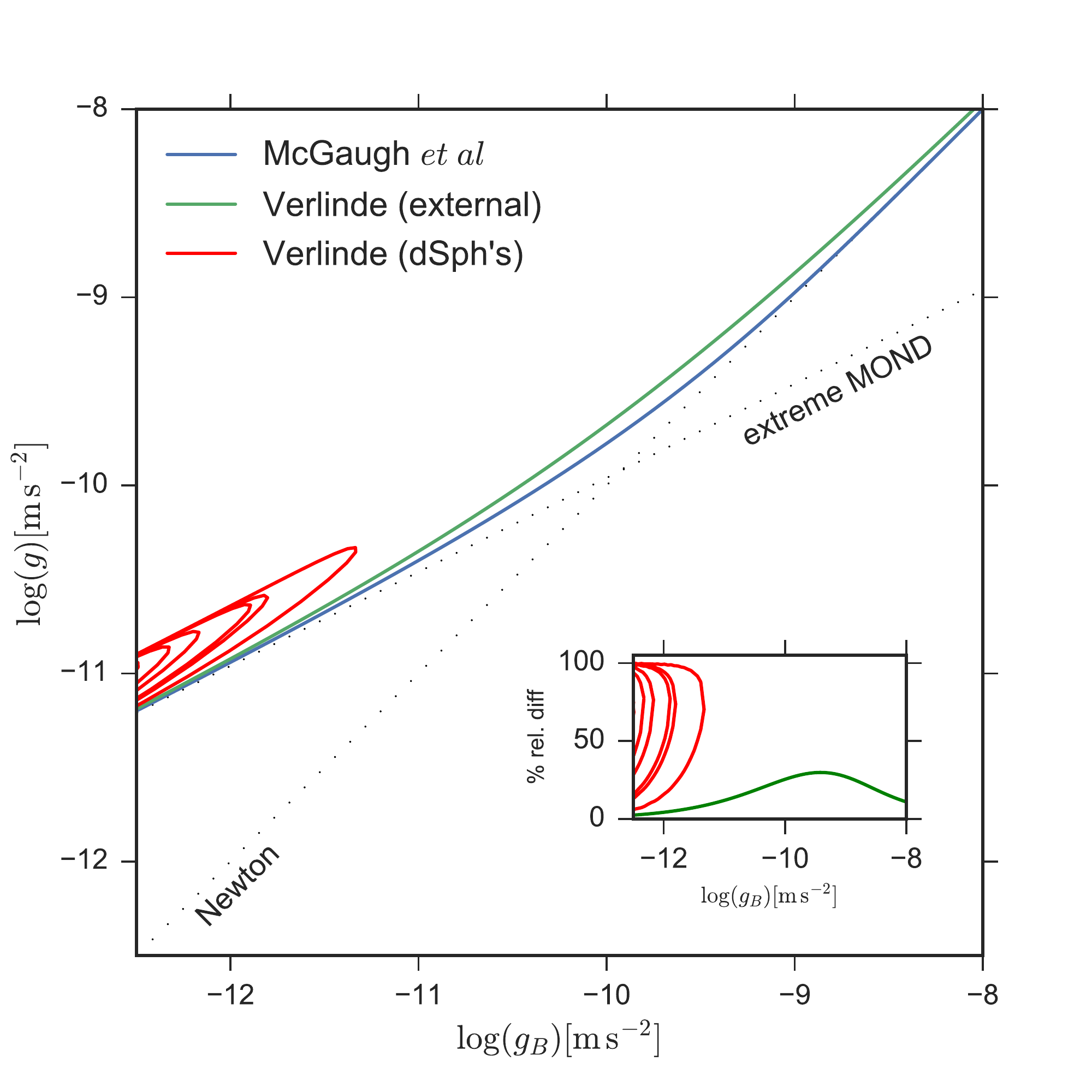}
\caption{Correlation between the baryonic acceleration, $g_B$, and the actually observed one, $g$, for different models. {\it {Blue line:} }empirical correlation as reported 
in~\citet{McGaugh:2016leg}. {\it Green line:} the prediction of Verlinde's theory outside a spherically symmetric massive object. At intermediate acceleration scales ($g_B\sim g\sim a_M$) Verlinde's model is marginally consistent with respect to the $20\%$ observational uncertainty of the McGaugh interpolating function, while both 
converge to the same limiting behaviour for large and small accelerations. {\it Red lines:} correlation for the classical eight dSph satellites of the Milky Way within the 
Verlinde's framework. We adopted a Plummer profile for the distribution of baryons in these objects, and an acceleration parameter of $a_0=5.4\times 10^{-10}\,\textrm{m/s}^2$,
consistent with current cosmological observations [see Eq.~(\ref{a0planck}) and the data of Table~\ref{tab:results}]. Note that the correlation between the two accelerations $g$ and $g_B$ depends, 
in general, on how baryons are distributed inside the configuration; a detailed description of this is shown in Figure~\ref{fig:mcVsVer2}. For reference, we add the 
corresponding correlation for the extreme MONDian (Milgrom's formula) and Newtonian regimes.}
\label{fig:mcVsVer}
\end{figure}

Following this line of thought, Mordehai Milgrom, in a series of seminal papers published in the {\it Astrophysical Journal} in 1983~\citep{Milgrom:1983ca,Milgrom:1983pn,Milgrom:1983zz}, proposed the original idea that 
the missing mass problem in galaxies could be resolved by a modification of Newton's law in the extremely weak field regime. In practice, this modified Newtonian dynamics (MOND) 
can be implemented in many different ways~\citep{Famaey:2011kh}. In this paper we use a non-conventional parametrization in terms of a phenomenological interpolating 
function $\zeta$, which relates the effective or observed gravitational field, $g$, to the standard Newtonian expression $g_B=GM_B/r^2$, in the following way:
\begin{equation}\label{eq.defzeta}
 g=\zeta(g_B/a_M)g_B\,.
\end{equation}
The function $M_B(r)$ denotes the mass distribution in the object, and the subscript $B$ in the different quantities stresses that they only contain baryons (in the cosmological acceptance 
of the word). Notice that there is a single free parameter in the model, a characteristic acceleration scale, $a_M$, which should be determined empirically. In our 
parametrization the interpolating function $\zeta$ depends only on $y_M\equiv g_B/a_M$. To reproduce Solar system observations we need that $\zeta(y_M\gg 1)=1$, 
thus $g=g_B$ in regions with large gravitational fields (when compared to the scale $a_M$). % if the gravitational acceleration is large when compared to the scale $a_M$.
If we further assume that $\zeta(y_M\ll 1)=1/\sqrt{y_M}$, the gravitational interaction will change from the usual $g=g_B$ in the Earth's laboratories, to the famous Milgrom's 
fitting formula $g=\sqrt{a_M g_B}$. This last expression holds in the extremely weak field regime, where $g_B\ll a_M$. 

It has been extensively argued in the literature that this simple modification of the gravitational interaction can reproduce the flattening of galactic rotational curves, as well as 
the baryonic Tully-Fisher relation in spiral galaxies, without postulating the existence of any new exotic matter component [see~\cite{Famaey:2011kh} for a recent 
review in the topic]. Apart from the conditions in the weak and strong field regimes described above, 
the function $\zeta(y_M)$ was not fully characterized in the original theory. However, in a recent study using more than 2500 data points in 153 rotationally supported 
galaxies of different morphologies and gas fractions,~\cite{McGaugh:2016leg} found a correlation between the observed and baryonic accelerations that can be 
interpreted in terms of the following interpolating MOND formula
\begin{equation}\label{mcgaugh}
 \zeta_{\rm McGaugh}(y_M)=\frac{1}{1-\exp(-\sqrt{y_M})}\,,
\end{equation}
with the characteristic acceleration given by $a_M= 1.20\pm 0.26\times 10^{-10}\, \textrm{m/s}^2$ [see also~\citet{Lelli:2017vgz} for more details on the analysis]. 
This function $\zeta_{\rm McGaugh}(y_M)$ naturally satisfies the two limiting 
conditions stated above for the cases of large ($y_M\gg1$) and low ($y_M\ll 1$) accelerations.

Even if Milgrom's MOND provides a satisfactory phenomenological description of the galactic kinematics, it exhibits tensions when trying to 
describe objects at different scales. For instance, galaxy observations suggest that $a_M\approx 1.2\times10^{-10}\, \textrm{m/s}^2$~\citep{1991MNRAS.249..523B}, consistent with the value reported 
in~\cite{McGaugh:2016leg}, whereas observations in galaxy clusters point to a value that differs by a factor of 3 or 4 \citep{Sanders:2002ue,Pointecouteau:2005mr}. 
Moreover, strong lensing in big galaxies point to an even larger discrepancy factor \citep{Famaey:2011kh}. These tensions may just signal that one should look at Milgrom's fitting formula as an effective description in the appropriate 
regime, expecting that the underlying modified theory of gravity will explain DM beyond the validity of MOND. 

Furthermore, there is an intriguing relation in this phenomenological model, where the characteristic acceleration scale used to explain the rotational curves in galaxies is of the same order of today's Hubble constant value~\citep{Milgrom:1998sy}. The link between dark energy (DE) and DM may be a hint for an emergent phenomenon of the 
underlying bricks of spacetime. Motivated by these arguments, Erik Verlinde has recently proposed that the emergent laws of gravity may contain an additional {\it dark} 
gravitational term that could replace DM in galaxy observations~\citep{Verlinde:2016toy}. From this perspective, what we usually identify as DM may just be the inescapable consequence of only having standard model particles in astrophysical objects.

The main result of this work is summarized in Figure~\ref{eq.defzeta}, where we show how the emergent gravity theory proposed by~\citet{Verlinde:2016toy} departs from 
the successful phenomenological MOND prescription, and in particular from the recently proposed interpolating function of~\cite{McGaugh:2016leg}. Bear in mind that the interpolating function of McGaugh {\it el al} is only one of 
the multiple possibilities to connect the MONDian and the usual Newtonian regimes in the region where the gravitational field is of the order of $a_M$. The main 
plot relates the observed or effective gravitational field $g$, to the standard Newtonian one $g_B$, inferred from the baryonic (stellar and gas) mass distribution. 
The blue line in Figure~\ref{eq.defzeta} denotes the fit to the observational data of McGaugh~{\it et al} based on the interpolating function in Eq.~(\ref{mcgaugh}), whereas the 
green one shows the prediction of Verlinde's model for the exterior of a spherically symmetric galaxy. Both descriptions agree at large and small accelerations, i.e. in the 
Newtonian and extreme MOND regimes, as expected. However, in the intermediate region they differ up to $30\%$, though this difference is still consistent with the 
observational uncertainties of the empirical relation of Eq.~(\ref{mcgaugh}). 

When it comes to the interior of galaxies, MOND-like theories (and then also the empirical relationship of McGaugh~{\it et al}) do not change their behaviour, staying along their 
corresponding lines of Figure~\ref{eq.defzeta}. However, for Verlinde's model the observed acceleration depends on gradients of the baryonic distribution of matter, increasing 
the observed gravitational field $g$ with respect the extreme MONDian regime. In red lines, one can appreciate the prediction of Verlinde's emergent gravity model for the classical 
eight dwarf spheroidal (dSph) galaxies of the Milky Way (enumerated in Table~\ref{tab:results}), resulting in a $100\%$ deviation from MOND well inside the galaxies.
Note that this deviation of Verlinde's model with respect to the predictions of MOND is consistent with  
current astrophysical observations, as has been identified in e.g.~\citet{Lelli:2017vgz}.  
Details of our results, and more about their implications, are discussed along the present work. 

The paper is organized as follows. In Section~\ref{sec.emergentgravity}, we outline the general picture developed in~\citet{Verlinde:2016toy}.
This section includes some of the technical details of Verlinde's work, which we expect to be short but comprehensive for non-experts in the subject. Among other things, we stress the 
motivation for an effective description of DM based on the theory of linear elasticity, highlighting the relevance of the maximal deformation assumption that 
was considered in the original work.
In Section~\ref{secMONDvsVerlinde}, we compare Verlinde's model with the successful phenomenological MOND description, and then argue in Section~\ref{sec.results} that, leaving
the maximal deformation assumption aside, Velinde's proposal can fit the velocity dispersions of the eight classical dSph satellites of the Milky Way with no further need of an extra dark particle component.
We finally present in Section~\ref{sec.discussion} a brief discussion on our findings.

\section{Verlinde's emergent gravity proposal}\label{sec.emergentgravity}

Classical General Relativity is a well-developed theory in Physics. However, we still lack consensus for its full quantum description, where a few candidates, such as string theory, have been put forward. Moreover, emergent 
phenomena, where collective behaviours arise from simple entities, occur in physical and other natural sciences all the time. The appearance of thermodynamics and 
hydrodynamics from microscopic states are probably two of the most notorious cases. There are many other well-known examples that illustrates this situation, such as
the Van der Waals force emerging from non-relativistic quantum electrodynamics~\citep{vanderwaals}, or the laws of classical mechanics from quantum theories when applied to 
large enough systems. 

Some theoreticians argue that there are two different categories of emergent phenomena~\citep{chalmers}: the weak and the strong one. 
In the former, it is the complexity of the system that makes it unfeasible to predict the collective behaviour, such as it happens with non-linear dynamics or classical 
phase transitions. In contrast, in the latter category, we find collective 
outcomes which in principle are not derived by the laws of their constituencies, even with a full knowledge of them. It is in this case where one may wonder if the laws of 
gravity, and perhaps even the structure of spacetime itself, unfold somehow as a collective result of microscopic degrees of freedom.

This is no longer a new idea. Early work dates back to more than 50 years, with the proposals of Wheeler~\citep{earlywork}, \cite{Finkelstein:1970xm}, and~\cite{Sakharov:1967pk}. However, in the last decade the topic has become more 
popular due to the particular connections between concepts such as gravitational entropy, entanglement, and the gauge/gravity duality. This last duality, in simple words, states that a 
field theory without gravity is equivalent to a higher dimensional theory containing gravity, supporting the idea of the gravitational field as an emergent concept. We refer 
the reader to the review papers by~\cite{Sindoni:2011ej}, \cite{Seiberg:2006wf}, \cite{Carlip:2012wa}, \cite{2014shst.book..213P}, and references within, where different approaches, challenges, and problems of emergent gravity are described 
in some detail. 

Among the vast constructions of emergent gravity, which usually break Lorentz invariance at the microscopic level~\citep{Sindoni:2011ej,Weinberg:1980kq}, there is one exception started by the work 
of Ted Jacobson ~\citep{Jacobson:1995ab}. In that work, he showed how Einstein equations can be recovered from the black hole entropy and the standard concepts of heat, entropy, 
and temperature. Given that thermodynamics reflects the collective behaviour of microstates, this particular perspective of gravity as an entropic force further 
supports the idea of an emergent spacetime. This early work was then further explored and generalized by Thanu Padmanabhan [see~\cite{Padmanabhan:2009vy}, and references therein]
and Erik Verlinde~\citep{Verlinde:2010hp}, highlighting the deep connection between General Relativity and entropy. A key argument in this relationship is the ``area law'' 
scaling of entropy, as opposed to the usual volume scaling~\citep{callen}. At the microscopic level, it is well understood that local interactions with a gapped Hamiltonian lead to an area 
law for the entanglement entropy~\citep{Eisert:2008ur}, connecting ideas in black hole physics (hence gravity), information theory, and quantum many body physics. 

Motivated by these arguments,~\cite{Verlinde:2016toy} has again extended this thermodynamical picture to explain the dark sector, containing the DM and 
DE components of our Universe. In the new proposal, he sketches without rigorous proofs how the (quasi)-de Sitter (dS) spacetime in which we presently live (the DE dominated
phase) can be obtained from a system of microstates, which are coherently excited above the true vacuum. The ground state or true vacuum corresponds to an anti-de Sitter (AdS) spacetime, which emerges from the fully entangled microstates. This particular construction may be better understood within the framework of the gauge/gravity correspondence, in 
terms of a particular representation of the microstates called the Multiscale Entanglement Renormalisation Ansatz (MERA) tensor network~\citep{mera}. 

A tensor network uses sites and connections to mathematically reduce the problem of finding the (vacuum) state in a system of many quantum particles. Using a coarse graining 
algorithm one can systematically reduce the number of degrees of freedom at each level. Then, after repeating it iteratively, one obtains a network. This mathematical description 
of many body physics has turned out to be very effective in studying condensed matter systems, and  has the interesting property of developing an effective metric 
along the network in many circumstances, see for example~\cite{Orus:2013kga}. In the case of the gauge/gravity duality, out of the possible tensor network descriptions, the particular choice of MERA develops 
an effective metric which results in a pure AdS when the continuum limit is taken~\citep{Nozaki:2012zj}. 

An important result is how the area law is obtained along this construction. The microstates are fully locally entangled, so that one can push the relevant information of the 
tensor network to the boundary of AdS. The relevant degrees of freedom to describe the full network are then those which live in the boundary of a lower dimensional space, 
which correspond precisely to the field theory in the duality~\citep{Ryu:2006bv}. Therefore, one does expect that the microstates (perhaps of this MERA tensor network) are the building 
blocks behind the AdS ground state.

From the AdS ground state, Verlinde argues that one could obtain a dS space (a vacuum spacetime with the DE component only), as an excitation %over this vacuum 
with very particular properties. One may expect a thermalized state due to the fact that dS has a horizon, and hence a Bekenstein's temperature associated to it. In order to achieve 
this homogeneous, very low energetic excited state, one may argue for long range correlations among the microstates, and hence not an area law but a volume scaling entropy.
This volume scaling entropy precisely matches the area law at the cosmological horizon, implying the entropy of DE within a spherical region of radius $r$ is 
\begin{equation}
 S_{DE}=\frac{r}{L} \frac{A(r)}{4G \hbar}\,.
\end{equation}
Here $L$ is the Hubble scale, related to the Hubble constant by $H_0=c/L$, and $A(r)$ is the area enclosing the spherical region. It was then conjectured by \cite{Verlinde:2016toy} that the DE entropy should be equally distributed 
among all the states, so that the excitations are delocalized and with a very slow dynamics, inhibiting all possible observation of DE in the lab. One may think that 
these non-local excitations that construct dS spacetime produce a very stiff structure, but on the contrary, the slow dynamics of these states makes it resemble more of an 
elastic medium. Given this resemblance, Verlinde really departs from the theory of elasticity to get predictions, since at the moment, there is no clear understanding of the real microstates describing the system. 

The idea of gravity being thought in terms of elasticity dates back to the work of~\citet{Sakharov:1967pk}, and more recently the possibility of describing dS in terms of an 
elastic medium was also worked out by~\cite{Padmanabhan:2004kf}, where he even argues how to obtain the value of the observed cosmological constant from 
this description. However, the real novelty in Verlinde's work is the study of how this dS scheme is affected by the intrusion of a baryonic mass $M_B$. On one hand, the 
cosmological horizon shrinks by the presence of the mass, reducing the entropy associated to the DE, as one can easily get convinced from the Schwarszchild-de-Sitter solution if the mass is 
placed near the origin at $r=0$ in the static patch of dS. On the other hand, this mass will carry a volume-scaling entropy given by $|S_M|=(2\pi M r)/\hbar$, as it can be shown 
by the effect of the gravitational potential produced due to the mass in the geodesic distance. If we then compare this mass entropy to the entropy lost by the medium by the presence
of the mass, we get
\begin{equation}\label{epsilondef}
 \epsilon(r)\equiv \frac{8\pi G}{a_0}\Sigma_B (r)\lessgtr 1\,,
\end{equation}
where $\Sigma_B\equiv M_B/A(r)$ is the surface density, and we have introduced the characteristic acceleration $a_0$ of the DE, given by $a_0=c\, H_0$. If $\epsilon>1$, more 
entropy is removed by the mass and the response to it is governed by the usual laws of Einstein's theory without dark fluids. If the opposite is true, $\epsilon<1$, then we 
are in the low surface density regime and there is a remnant of the DE entropy in the volume occupied by the mass, which behaves as an incompressible elastic medium and will 
affect the gravitational laws associated with the baryonic mass, mimicking a DM component. 

It turns out that $\epsilon$ corresponds to the largest principal strain of the elastic medium, namely $(\epsilon_{ij}-\epsilon_{kk}\delta_{ij})n_{j}=\epsilon n_{i}$, where 
$\epsilon_{ij}$ is the strain tensor and $n_i$ is the maximal strain (unitary) direction. Therefore, the response of the medium to the mass inclusion can be fully understood 
in terms of the stress and the strain of the elastic theory. The linear theory of elasticity considered by~\citet{Verlinde:2016toy} is such that no pressure waves are present and the 
displacement is produced by a scalar quantity only, namely, the stress tensor $\sigma_{ij}$ obeys $\sigma_{ij}=a_0^2 (\epsilon_{ij}-\epsilon_{kk}\delta_{ij})/(8\pi G)$. This 
corresponds to a theory of gravity with a central force, described by a scalar potential $\Phi$ in terms of the elastic medium displacement $u_i$, and given by $\Phi=a_0 u_i n_i$.

In order to obtain the main result of~\citet{Verlinde:2016toy}, Eq.~(\ref{epsilondef}) can be thought of in a different way. If we are in the dark sector regime, $\epsilon<1$, the 
observed surface density produced by baryons is given by $\Sigma_{D}\equiv a_0\epsilon/8\pi G$. From the elastic energy due to the inclusion of a mass $M_B$, Verlinde 
arrives to the key equation
\begin{equation}\label{generalformula}
\left(\frac{8\pi G}{a_0}\Sigma_D\right)^2\leq - \frac{2}{3}\nabla_{i}\left(\frac{\Phi_B}{a_0}n_i\right)\,,
\end{equation}
where the equality is only achieved when the largest principal strain $\epsilon$ takes its maximal value (so the perpendicular strains are all equal), and the medium response is 
negligible well outside the mass. In other words, observations may only put a lower bound on the parameter $a_0$, since a larger value can be accommodated by having a smaller 
elongation (or compression) of the elastic medium due to the baryonic mass inclusion.

In the case of a maximal value for the largest principal strain, and further assuming spherical symmetry, we have that $\Sigma_B=-\Phi_B/(4\pi G\, r)=M_B(r)/A(r)$, where 
$M_B(r)=\int_0^r \nu(y)A(y)dy$, and $\nu(r)$ is the mass density function of baryons. Then, the previous key formula of Eq.~(\ref{generalformula}) reduces to our starting point for the analysis in this paper, namely
\begin{equation}\label{eq.MD}
 M^2_D(r)=\frac{a_0 r^2}{6G}\frac{d}{dr}\left(rM_B(r)\right)\,.
\end{equation}
As stated earlier, the parameter $a_0=cH_0$ is a characteristic acceleration scale in the theory related to the radius of the Universe. %(quasi)-dS universe we seem to inhabit. 
However, the previous result holds while assuming matter is a small perturbation over a dS background, which is not entirely true in the present Universe. % in today's matter distribution. 
Since Verlinde's proposal does not provide a cosmological evolution of fields, one cannot extract the value of $a_0$ from a time-dependent Hubble parameter. However, if 
one assumes DE is a cosmological constant and the cosmic evolution resembles roughly that of General Relativity, in the distant 
future the Hubble parameter would remain practically constant while the matter structures would really be small perturbations around a dS background, 
justifying the static derivation of Verlinde's formula. Under this assumption, we decide to use today's DE density as the parameter $a_0$ in the model.
%If the DE continues to dominate we should reach a state that is closer to Verlinde's working assumption. Moreover, it is strictly impossible to decide the precise value of $a_0$ from first principles, given that the original elastic medium proposal does not contain time evolution nor is being formally derived from a theory of microstates. In order to choose a theoretical value for $a_0$ in our analysis, we decide to use a de Sitter scale matched to the present dark energy density, which should be preserved over time if the DE density does not evolve, as it happens with a cosmological constant. 
Therefore, the value of $H_0$ is not just the value of the Hubble parameter today, which has other contributions apart from the DE.
Using the analysis of current cosmological observations by~\citet{Ade:2015xua},\footnote{We use the cosmological parameters as reported in~\citet{Ade:2015xua}, obtained from
an analysis of the TT, TE, EE, and LowP datasets, together with their 68\% confidence limits as errors.}
we obtain $a_0= 5.41\pm 0.06\times10^{-10}\,\textrm{m/s}^2$. Based on this observational value, we fix our theoretical input of the dS acceleration scale to be
\begin{equation}\label{a0planck}
a_0=5.4\times10^{-10}\,\textrm{m/s}^2.
\end{equation}
Note that the error contribution of this approximation is negligible when compared to other systematics. Furthermore, given the arbitrariness in the original proposal to restrict
$a_0$ to a precise value, we also perform a secondary test on the data without fixing the value of $a_0$ in Section~\ref{sec.results}.
%, and the value remains compatible with other observations, such as those in~\cite{1991MNRAS.249..523B} and~\cite{McGaugh:2016leg}, as we clarify below.

\begin{figure}
\includegraphics[width=\columnwidth]{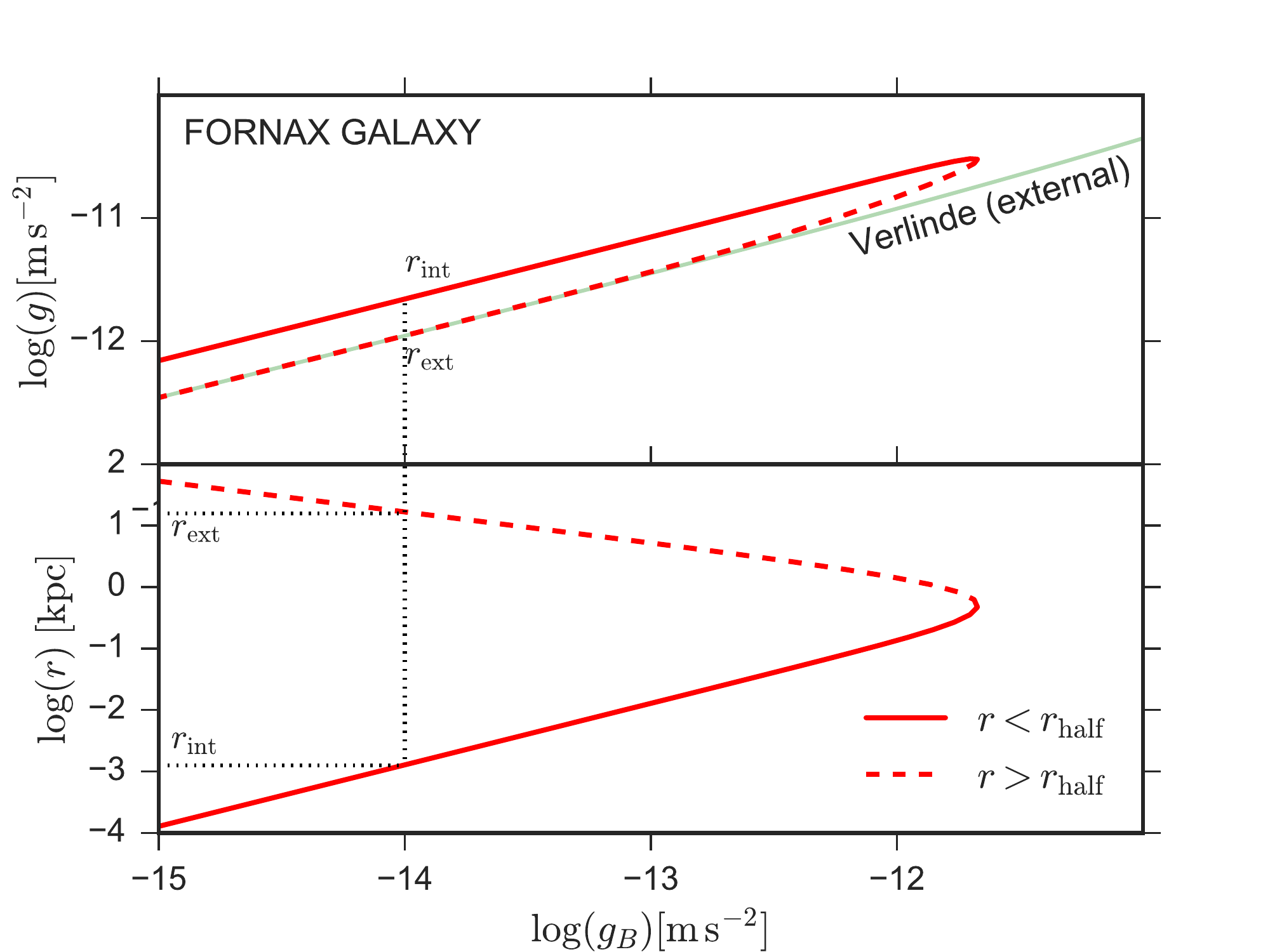} 
\caption{Detailed behaviour of the effective gravitational acceleration, $g$, as a function of the baryonic component, $g_B$, within Verlinde's theoretical framework. 
This relation depends on whether we are exploring an inner or an outer galactic region, as shown by the same baryonic acceleration at two different radii, $r_{\rm ext}$ 
and $r_{\rm int}$. We use a Plummer density profile to model the stellar mass distribution, and the structural parameters of
Table~\ref{tab:results} for the Fornax galaxy (see the red lines in Figure~\ref{fig:mcVsVer} for the other classical dSph's).
The characteristic radius that separates the inner and outer regions is close to the half-light radius, but they do not coincide exactly.  
The maximum difference with respect to the external profile is of $100\%$, and it is independent of the galaxy's details, as long as it is regular at the origin. 
This inner region behaviour is expected in general, since the acceleration in Verlinde's theory depends 
on the derivatives of the baryonic mass profile; see the text for a detailed explanation.}
\label{fig:mcVsVer2}
\end{figure}

\section{Verlinde's emergent gravity versus MOND}\label{secMONDvsVerlinde}

Baryons and the DM phenomena are highly correlated in this new gravitational picture, and this is a prediction that must be contrasted against observations.
For a point particle, or outside and extended spherically symmetric object, the mass in baryons remains constant. From 
Eq.~(\ref{eq.MD}), $M_B(r)=\textrm{const.}$ translates to $M_D^2(r)=a_0M_Br^2/6G$.
In the extremely weak field regime, when $g_B\ll a_0$, we recover Milgrom's original fitting formula $g(r)=\sqrt{a_Mg_B(r)}$, where $g_B=G M_B/r^2$. 
%Note that the factor $a_M=a_0/6$ is now related in an obvious way to cosmological quantities, 
Note that the characteristic scale $a_M$ is now related in an obvious way to cosmological quantities, $a_M=a_0/6$, and that the value of $a_0$ in Eq.~(\ref{a0planck}) 
remains compatible with %that of $a_M$ inferred 
previous estimations by~\cite{1991MNRAS.249..523B} and~\cite{McGaugh:2016leg} using galactic observations. This seems to suggest that one can explain the kinematics of spiral galaxies as well as 
MOND can do. However, that is not entirely true at this level, since  the  derivation of Eq.~(\ref{eq.MD}) was based on spherical symmetry, which is not fulfilled by baryons in 
those objects. Furthermore, Milgrom's fitting formula is only recovered in the exterior of a massive object, but the new fitting formula differs inside matter distributions, 
as we discuss in detail in what follows. We should then test the viability of this new expression using astrophysical observations which are in agreement with the hypothesis behind the deriviation of Eq.~(\ref{eq.MD}).  
%For recent work on the  observability of Verlinde's model see~\cite{Liu:2016nwt}, \cite{Brouwer:2016dvq}, and~\cite{Iorio:2016qzr}.

One may wonder how Verlinde's model differs from MOND for spherical objects. At this point it is interesting to stress that, for the emergent gravity model outlined in Section~\ref{sec.emergentgravity}, it is still possible to relate the baryonic acceleration, $g_B$, to the actual observed one, $g$, through an interpolating function of the form 
\begin{equation}\label{eq.zeta.verlinde}
\zeta_{\rm Verlinde}(y,y')=1+\sqrt{\frac{3y+ry'}{6y^2}}
\,.\end{equation}
Contrary to what happens in the case of MOND, see Eq.~(\ref{eq.defzeta}) for reference, this function depends not only on $y=g_B/a_0$ (i.e. $y_M=g_B/a_M=6y$), but also on its radial derivative, 
$y'=dy/dr$. This derivative dependence of the interpolating function is expected to hold away from spherical symmetry, as one may appreciate from the general expression in 
Eq.~(\ref{generalformula}), and it represents the stronger deviation from the MOND paradigm. This effect has already been pointed out by~\citet{Verlinde:2016toy}, where he 
argues that the slope of a density profile can explain the orders of magnitude in cluster observations, or even how weak lensing systems could be affected. See also the work 
in~\cite{Liu:2016nwt}, \cite{Brouwer:2016dvq}, and~\cite{Iorio:2016qzr}. In this paper we explore the inner part of galaxies in more detail, and in particular for those which are almost spherical in symmetry. 

Outside an extended spherically symmetric massive object we have that $ry'=-2y$, hence an effective MOND-like regime with $\zeta_{\rm Verlinde}^{\rm out}(y)=1+1/\sqrt{6y}$ 
naturally emerges as a particular limit of Verlinde's theory. % and when the matter distribution plays no significant role. 
Note that once we identify $y_M=6y$, the two limiting cases %($y\ll 1$ and $y\gg 1$) 
match the Newtonian, $\zeta_{\rm Verlinde}^{\rm out}(y\gg 1)=1$, and the extreme MONDian, $\zeta_{\rm Verlinde}^{\rm out}(y\ll 1)=1/\sqrt{y_M}$, regimes in the exterior region. % i.e. $a_M=a_0/6$. 

While still considering the exterior of a massive object, one may wonder how the interpolating function in Eq.~(\ref{mcgaugh}) compares to the Verlinde's model away from the 
extreme MOND regime. In Figure~\ref{fig:mcVsVer} we show the correlation between the two acceleration profiles $g$ and $g_B$, using the same scales as in Figure~3 of~\cite{McGaugh:2016leg}. 
The blue line shows the result of the analysis carried out by McGaugh~{\it et al}, a fit provided by 2693 data points in 153 rotationally 
supported objects of the SPARC database. The green line corresponds to the MOND-like regime that emerges outside the exterior of a spherically symmetric 
massive object in Verlinde's theory. Note that the relative error of Verlinde's model with respect to the McGaugh's prediction %between these two different models 
is never larger than $30\%$, marginally accepted by the observations~\citep{McGaugh:2016leg}, and both descriptions coincide in the Newtonian and the extreme MOND 
limits, as previously stated.

On the contrary, for the interior region of the object, the $y'$ dependence in the interpolating function implies all relevant expressions cannot be independent of the density profile.
In other words, the relation between $g$ and $g_B$ depends, in general, on the Newtonian potential profile, that is, on how baryons are distributed in the configuration.
It is important to stress that this feature makes it possible to distinguish, at least in principle, Verlinde's emergent gravity from MOND when using peculiar velocities in the 
inner galactic regions.

In MOND, the interior and the exterior regions describe the same acceleration profile, see the blue line in Figure~\ref{fig:mcVsVer}. However, the radial derivative 
dependence of the interpolating function in Verlinde's model gets the interior profile of $g$ to depart drastically from the outer region. Although one must select a 
particular baryonic configuration model to describe in more detail the differences between the exterior and interior regions, some general features can still be drawn for the case of a regular self-gravitating object. In order to appreciate this, note that the baryonic mass scales as $M_B(r)\sim r^3$ in the innermost region of a regular mass configuration, hence $y$ scales as %we can approximate %we can conclude that 
$ry'=y$ close to the centre of these objects. By introducing this expression in Eq.~(\ref{eq.zeta.verlinde}) we obtain that the baryonic, $g_B$, and the observed, $g$, gravitational accelerations can be related through 
the expression $\zeta_{\rm Verlinde}^{\rm in}(y\gg 1)=1/\sqrt{y_M/4}$. This corresponds to an effective MOND parameter that has been renormalized by a factor of 4 
with respect to its value outside the configuration, explaining the 100\% departure from the MONDian regime in Figure~\ref{fig:mcVsVer}. 

In order to see how these two generic regimes, the close to the centre and the exterior one, connect in the intermediate region of the mass configuration, one needs 
to propose a concrete profile for the baryonic distribution. One of particular interest is the Plummer density profile~\citep{1911MNRAS..71..460P}, since it has been 
extensively used in the literature to describe (nearly) spherically symmetric stellar structures such as globular clusters, galactic bulges, or dwarf spheroidal 
galaxies (see for instance Chapters~4 and~7 in~\cite{Binney} for details on the profile and its applications). Current observations seem to indicate that globular 
clusters or galactic bulges are almost DM free~\citep{1996IAUS..174..303H,2013MNRAS.428.3648I,2017MNRAS.465.1621P}, while dwarf spheroidals (dSph's) are expected to 
be dominated by the dark component, as discussed in~\cite{2013pss5.book.1039W}, hence our interest in dSph's as a probe of Verlinde's model. 

The Plummer mass density profile is given by
\begin{equation}\label{eq.3density}
 \nu(r)=\frac{3\Upsilon_* L}{4\pi r _{\textrm{half}}^3}\frac{1}{[1+(r/r_{\textrm{half}})^2]^{5/2}}\,,
\end{equation}
where $\Upsilon_*=M_*/L$ is the stellar mass-to-light ratio, $M_*$ and $L$ are the stellar mass and luminosity of the object, respectively, and $r_{\textrm{half}}$ 
is the half-light radius, i.e. that radius enclosing half of the total luminosity. We will give more details about this profile and the dSph's in the following 
section, but for the moment, we just consider it as an example to explain the behaviour of the acceleration in the inner and outer regions, and its relation to the 
baryonic acceleration for an extended stellar configuration. The result is illustrated in Figure~\ref{fig:mcVsVer2}, for the particular case of one dSph: Fornax. 
A turning point is precisely the expected behaviour for a $g$ versus $g_B$ diagram, where an up-then-down density profile always leads to a forward-then-backwards 
acceleration function. Verlinde's model is different to MOND since the radial derivative in~(\ref{eq.MD}) moves the turning point away from the outer region 
behaviour. The discrepancy of Verlinde's emergent gravity with respect to MOND in the interior field rises to $100\%$, always to regions of larger $g$, well inside 
the galaxy, as it was previously anticipated when discussing the renormalized effective MOND parameter that emerges close to the central region of a regular object. 
The red lines in Figure~\ref{fig:mcVsVer} show the prediction for the eight classical dSph satellites of the Milky Way using the Plummer profile with the parameters 
reported in Table~\ref{tab:results}. Note that this behaviour away from the extreme MONDian regime is consistent with the scatter of the dSph's away from the 
interpolating function of Eq.~(\ref{mcgaugh}), as has been shown in the recent observational study of~\cite{Lelli:2017vgz}.

\begin{table*}
\begin{center}
\scriptsize
\begin{tabular}{lcccccccccc|cccccc}
\\\hline\hline
\multicolumn{8}{c}{} &  \multicolumn{3}{c}{Test 1 ($a_0=a_{0,\textrm{theory}}$)}& &\multicolumn{5}{c}{Test 2 ($a_0$ unspecified)} \\
%\cline{9-11} \cline{13-17}
\cline{8-17}
$\;$Object$\quad$ & & $L_V$ & & $r_{\textrm{half}}$ & & $\langle\sigma_{\textrm{los}}\rangle$ & & $\Upsilon_*$ & $\beta$&$g/g_{\rm MW}$& & $a_0$ &   & $\Upsilon_*$ & & $\beta$ \\ 
&& $(L_{V,\odot})$ & & $(\textrm{pc})$ & & $(\textrm{km\,s}^{-1})$ & & $(M_{\odot}\,L^{-1}_{V,\odot})$ & & & &$(\textrm{m\,s}^{-2})$ & &$(M_{\odot}\,L^{-1}_{V,\odot})$& \\\hline 
$\;$Fornax  & & $2.0\times 10^{7}$ & & $710 \pm 77 $ & & $11.7\pm 0.9$  & & $0.32$ & $-0.15$& $0.15$& & $1.5\times 10^{-10}$ & & $1.26$ & & $-0.15$\\
$\;$Sculptor & & $2.3\times 10^{6}$ & &$283\pm 45$ & & $9.2\pm 1.1$ & & $1.58$ & $-1$& $0.06$& &$7.5\times 10^{-10}$ & & $1.35$ & & $-0.98$ \\
$\;$Carina  &  & $3.8\times 10^{5}$ & & $250\pm 39$ & & $6.6\pm 1.2$  & &  $2.23$ & $-0.97$&$0.07$ & &$9.5\times 10^{-10}$ & & $1.4$ & & $-0.96$\\
$\;$Draco  & & $2.9\times 10^{5}$ & & $221\pm19$ & & $9.1\pm 1.2$ &  & $13.08$ & $<-1.5$&$0.04$ & &$2.7\times 10^{-9}$ & & $3.2$ & & $<-0.57$ \\
$\;$Leo I  & & $5.5\times 10^{6}$ & & $251\pm27$ & & $9.2\pm 1.4$ &  & $8.69$ & $-1$&$0.004$ & & $2.2\times 10^{-9}$ & & $2.63$ & & $-1.08$ \\
$\;$Leo II  & & $7.4\times 10^{5}$ & & $176\pm42$ & & $6.6\pm 0.7$  & & $2.38$ & $0.13$&$0.01$ & &$1.0\times 10^{-9}$ & & $1.48$ & & $0.12$\\
$\;$Sextans  & &  $4.4\times 10^{5}$ & & $695\pm44$ & & $7.9\pm 1.3$  & & $2.09$ & $-0.21$& $0.63$ & & $9.5\times 10^{-10}$ & & $1.22$ & & $-0.21$\\
$\;$Ursa Minor  & &  $2.9\times 10^{5}$ & & $181\pm27$ & & $9.5\pm 1.2$  & & $11.31$ & $-1.8$&$0.03$ & & $2.6\times 10^{-9}$ & & $3.0$ & & $-1.83$\\
\hline
\end{tabular}
\caption{Luminosity in the $V$-band $L_V$, half-light radius $r_{\textrm{half}}$, and mean velocities $\langle\sigma_{\textrm{los}}\rangle$, for the different Milky Way's 
dSph galaxies, as reported in~\citet{2012AJ....144....4M}. {\it Test 1:}  Median values of the marginalized posterior distribution for the stellar mass-to-light ratio $\Upsilon_*$, and orbital 
anisotropy $\beta$, taking the acceleration scale fixed to the theoretical value in Eq.~(\ref{a0planck}). {\it Test 2:} Median values of the marginalized 
posterior distribution for the acceleration scale $a_0$, stellar mass-to-light ratio $\Upsilon_*$, and orbital anisotropy $\beta$. The theoretical value of $a_0$ is consistent 
at 2$\sigma$ level for all galaxies. Moreover, there is a strong degeneracy between $a_0$ and $\Upsilon_*$, which may drive $\Upsilon_*$ to a lower, observationally accepted value at the cost of 
increasing $a_0$, which is consistent with not knowing the actual deformation profile of the elastic background medium. 
Note that the values of $\beta$ in both tests are consistent, pointing to no degeneracy with the other parameters.}
\label{tab:results}
\end{center}
\end{table*}

\section{Milky Way's Dwarf Spheroidals in the Emergent Gravity picture}\label{sec.results}

Given the current degree of development of the theory, and until we can generalize the expression in Eq.~(\ref{eq.MD}) to configurations with less symmetry, nearby dSph 
galaxies are probably one of the most promising objects to test Verlinde's new proposal. These galaxies are the smallest and least luminous in the Local Group, and given 
their proximity to us they are relatively well understood. Stellar population studies point to a negligible gas contribution and stellar mass-to-light ratios (using the Sun as our reference unit) in the range 
$1\lesssim \Upsilon_*\lesssim 3$~\citep{Bell:2000jt}. However, there is evidence that they may require an absolute (not only stellar) mass-to-light ratio as large as 
$\Upsilon\sim 10-10^{2}$ in order to explain their internal kinematics~\citep{2013pss5.book.1039W}. The key point is that these systems reach the regime where, if we assume there is no DM component, standard 
Newtonian theory completely fails and %, a point where 
we expect the heart of Verlinde's proposal to be revealed.

The Plummer density profile of Eq.~(\ref{eq.3density}) provides a good description of the baryonic distribution in the eight classical dSph satellites of the Milky Way, 
with the possible exceptions of Fornax and Leo. 
 These two galaxies are not perfectly fitted by the Plummer profile in the outer regions, presumably because they have not been affected by tidal disruptions~\citep{Penarrubia:2008mu}.
Nevertheless, we use the Plummer profile for all the galaxies in the sample in order to compare with other analysis in the literature.
%It is believed that the absence of stellar tidal disruptions by the Milky Way is the cause of not having a power-law outer region, consistent with the Plummer profile, 
%for these two particular dPhs~\citep{Penarrubia:2008mu}. 
We can read their structural parameters from Tables~4 and~5 of~\cite{2012AJ....144....4M}, which, for convenience, we also show in our Table~\ref{tab:results}.

The internal coherent rotation in these dSph's is negligible, and the stellar component is supported against gravity by its random motion. Contrary to the case of spiral galaxies, 
the observable that can be used is not a rotation curve but, rather, the line-of-sight velocity dispersion of its components. In particular, in this work we use the data of~\cite{Walker:2009zp,Walker:2007ju}
for the internal dynamics of the eight classical dSph  satellites  of  the  Milky  Way. 
We assume, as usual, that each galaxy is spherically symmetric and in dynamical equilibrium. The stars trace the underlying 
gravitational potential, in this case determined by the effective mass formula $M(r)=M_B(r)+M_D(r)$, with $M_D(r)$ given in Eq.~(\ref{eq.MD}).
We also consider that this dynamical equilibrium approximates well a static mass configuration, which together with the spherical symmetry reflects Verlinde's key assumptions.

An analysis of the velocity dispersion profiles shows that the quantity that is best constrained in a dSph galaxy is the enclosed mass at the half-light radius, $r=r_{\rm half}$~\citep{Walker:2007ju}. 
To qualitatively compare MOND and Verlinde's proposal, we search for the scaling factor needed in the MOND's mass-to-light ratio to match Verlinde's profile at the 
half-light radius. If baryons follow a Plummer distribution, we find that this scaling factor is $\Upsilon^{\rm M}_*=5/2 \Upsilon_*$, see Appendix~A for details. This 
simple result means that MOND may require a mass-to-light ratio more than two times larger than Verlinde's one for the same value of $a_0$. It is important to mention that the stellar mass-to-light ratios in these dSph's are unknown. A study by~\cite{Kirby:2013wna} based on stellar formation histories (and independent from kinematics) finds a mean value of $\Upsilon_*=1.03$, while a mean of  $\Upsilon_*=1.6$ 
was previously found by~\cite{Woo:2008gg}. In general, it seems that stellar population studies point to values not much larger than unity~\citep{Bell:2000jt}, which may 
suggest Verlinde's model is favoured along this line of thought, given the larger values that seem to be preferred by MOND according to a study carried out 
by~\cite{Angus:2008vs}. 
However, a detailed analysis of Verlinde's model should be performed using each dSph's data, which is the goal of 
the rest of this section.

In order to do a more quantitative analysis of the Verlinde's model and how well it describes dSph's kinematics, we perform a standard parametric Jeans analysis as was 
done in~\citep{Walker:2009zp,Walker:2007ju}. See the Section~4.8 in~\cite{Binney} for a comprehensive review on these techniques. The (observed) line-of-sight velocity dispersion of the dSph's, 
$\sigma_{\textrm{los}}^2(R)$, can be related to the (modelled) mass profile, $M(r)$, and stellar mass density, $\nu(R)$, through
\begin{eqnarray}
 \sigma_{\textrm{los}}^2(R) &=& \frac{2G}{I(R)}\int_R^{\infty}dr'\nu(r')M(r')(r')^{2\beta-2}\times\nonumber\\
 && \int_R^{r'} dr\left(1-\beta\frac{R^2}{r^2} \right)\frac{r^{-2\beta+1}}{\sqrt{r^2-R^2}}\,,
 \label{eq:sigmalos}
\end{eqnarray}
where $R$ is the projected radius, and $\beta(r)$ represents the orbital anisotropy. This anisotropy is not constrained observationally and, in general, could depend on the radius. However, as it is customary in a first study, we assume it to be a constant in this work. 
The stellar mass density profile in Eq.~(\ref{eq.3density}) is recovered from the observed 2-dimensional (projected along the line-of-sight) stellar density  
\begin{equation}
 I(R)=\frac{L}{\pi r_{\textrm{half}}^2}\frac{1}{[1+(R/r_{\textrm{half}})^2]^2}\,.
\end{equation}

In order to fit the observational data we have three free parameters per galaxy.
%We fit the observational data using the three free parameters per galaxy; 
One associated with the theory: the acceleration scale $a_0$; and two more related to the stellar component: the stellar mass-to-light ratio $\Upsilon_*$ and the orbital anisotropy $\beta$. Note that this differs from a standard DM particle analysis, where the mass-to-light ratio is uncorrelated to the DM density profile, and the luminosity cancels out in Eq.~(\ref{eq:sigmalos}). To proceed further, we perform a Markov Chain 
Monte Carlo (MCMC) analysis to explore the parameter space and estimate the median values of the different quantities (see Appendix~\ref{appB} for details on 
the MCMC analysis).
Since the acceleration scale is a constant in the theory, one should perform a joint analysis for the eight galaxies keeping $a_0$ fixed. However, as we mentioned earlier, the 
parameter $a_0$ may be thought as encoding the hidden information about the elastic medium deformation if the principal strain does not take its maximal value, as it was assumed 
when taken the equality in Eq.~(\ref{generalformula}) to deduce expression~(\ref{eq.MD}). 
Therefore, we perform two different tests that, steep by steep, help us to better understand the behaviour of the model when trying to fit the internal kinematics of the 
eight classical dSph's:

{\it Test 1:} Fix the value of the acceleration scale to the prediction of the theory, $a_0=5.4\times 10^{-10}\,\textrm{m/s}^2$, 
and look for the most probable values of the  orbital anisotropy $\beta$, and stellar mass-to-light ratio $\Upsilon_*$, compatible with the data of each individual object. For this test we assume uniform priors 
in the following range
\begin{subequations}\label{eq:prior1}
\begin{eqnarray}
 -2 <  &\ln(\Upsilon_*[\Upsilon_{\odot}])& < 5\,, \\
 -3 <  &-\ln (1-\beta)& < 3\,.
\end{eqnarray}
\end{subequations}
The main purpose of this test is to establish if the theory, as it reads in Eq.~(\ref{eq.MD}), needs to include any extra DM, assuming the rest of Verlinde's hypotheses 
are satisfied. 

\begin{figure*}
 \includegraphics[width=\columnwidth]{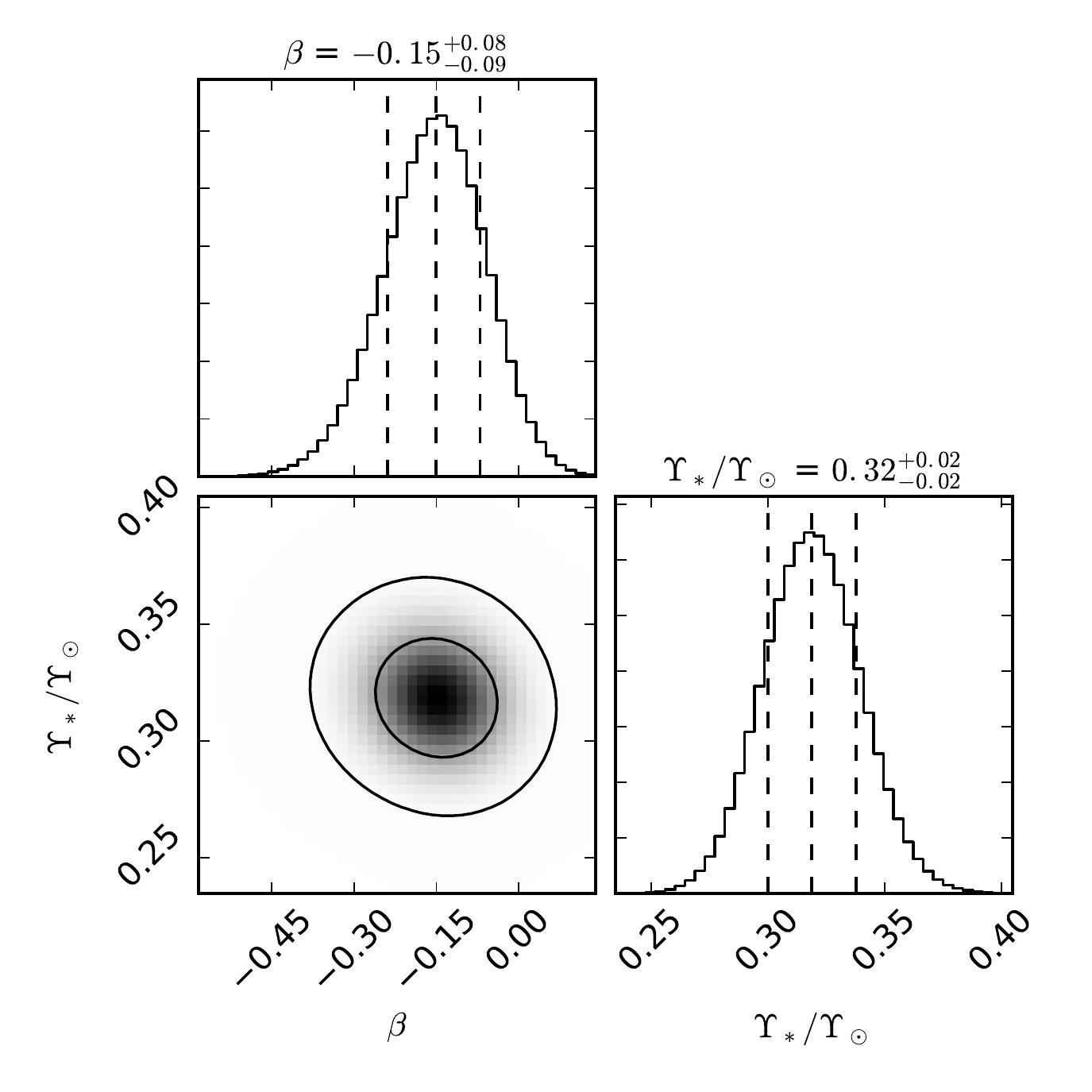}  
 \includegraphics[width=\columnwidth]{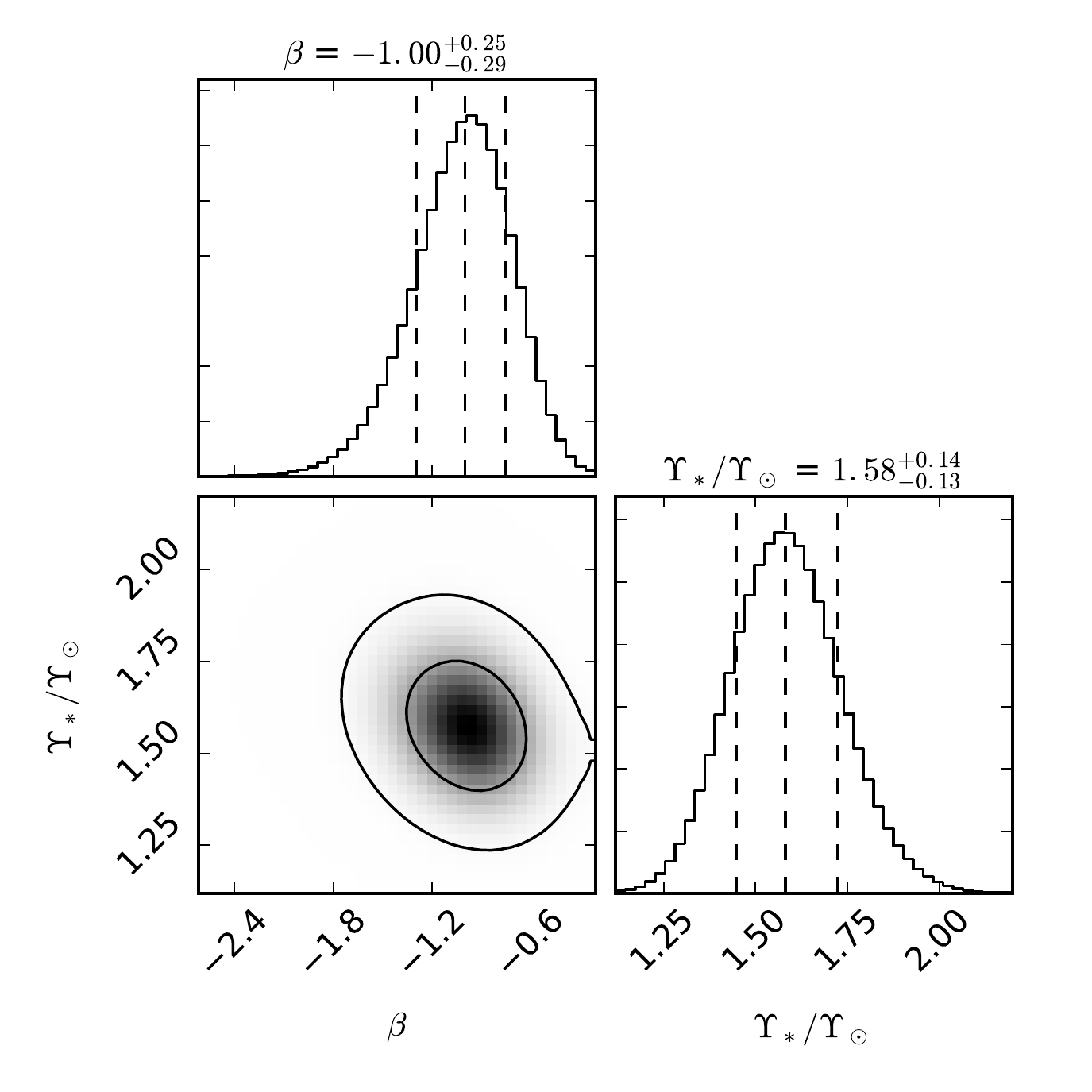}
 \caption{{\it Test 1:} Posterior distribution function of the anisotropy and stellar mass-to-light ratio in Fornax (left) and Sculptor (right), 
 resulting from the fit to their velocity dispersion profile under Verlinde's theory, assuming a constant $a_0=5.4\times 10^{-10}\, {\rm m/s}^{-2}$. Both parameters are well 
 constrained and show no degeneracy between them.  The other galaxies show similar results.
 }
 \label{fig:test1}
\end{figure*}

The results of this test are summarized in Table~\ref{tab:results}. Note that %within this test, 
there is some tension between the observationally allowed range of 
values for the stellar mass-to-light ratios, expected to be of the order of 1, as previously argued, and those in Draco, $\Upsilon_*=13.08$, Ursa Minor, $\Upsilon_*=11.31$, and Leo~I, $\Upsilon_*=8.69$. Furthermore, the value of the mass-to-light 
ratio for Fornax is quite low as well, $\Upsilon_*=0.32$. In Figure~\ref{fig:test1} we show, for illustration, the posterior distribution for Fornax and Sculptor of both parameters:
the anisotropy and the stellar mass-to-light ratio. Similar posterior distributions appear for the other objects in the sample. Note they are well constrained and show no degeneracies. 

{\it Test 2:} Find the most probable values of the acceleration scale $a_0$, orbital anisotropy $\beta$, and stellar mass-to-light ratio $\Upsilon_*$, 
compatible with the data of each individual object. In this case, we adopt uniform priors in the following range 
\begin{subequations}\label{eq:prior2}
\begin{eqnarray}
 -12 < & \log(a_0[\textrm{m\,s}^{-2}]) & < -8\,, \\
 -2 <  &\ln(\Upsilon_*[\Upsilon_{\odot}])& < 2.5\,, \\
 -3 <  &-\ln (1-\beta)& < 3\,.
\end{eqnarray}
\end{subequations}
The goal of this test is to explore if by relaxing the assumption of a maximal deformation, the analysis drives the stellar mass-to-light ratio to values in the range 
$1\lesssim\Upsilon_*\lesssim3$, consistent with the current stellar population and formation history studies of these objects. Remember that, according to the inequality (\ref{generalformula}), a maximal deformation can be interpreted 
as the scaling of the acceleration $a_0$ to higher values from its theoretical value.

\begin{figure*}
 \includegraphics[width=\columnwidth]{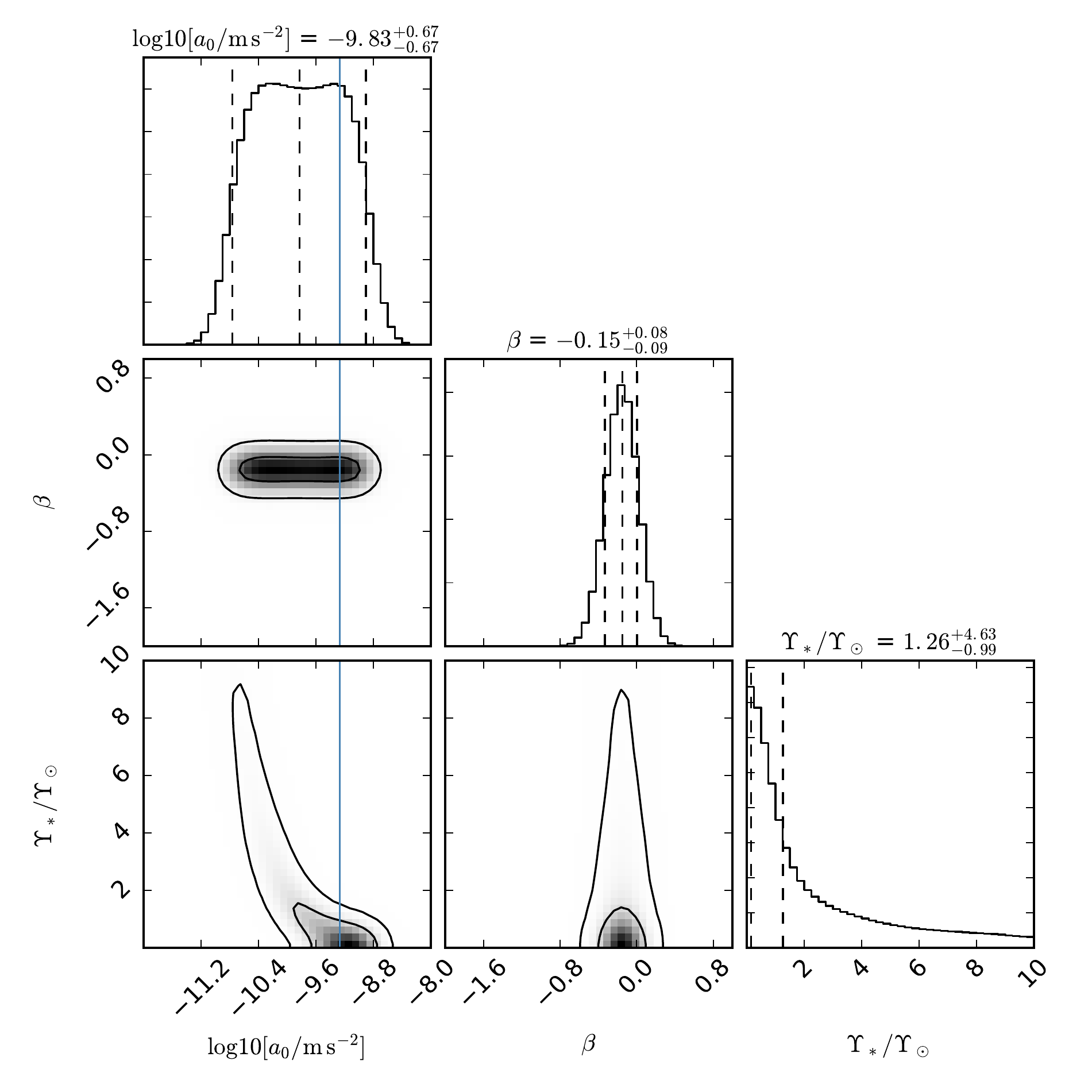}  
 \includegraphics[width=\columnwidth]{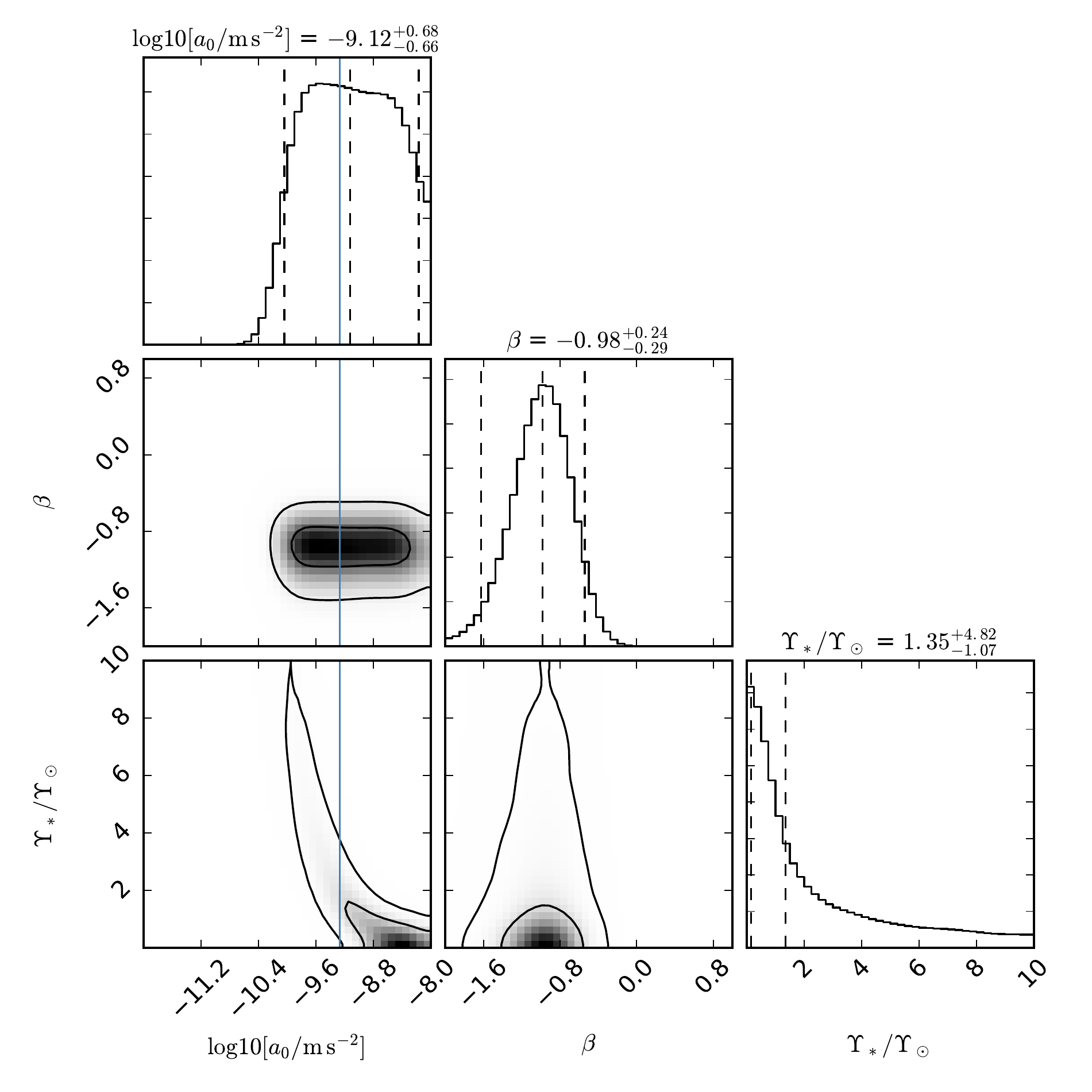}
 \caption{{\it Test 2:} Posterior distribution function of the anisotropy, stellar mass-to-light ratio, and acceleration scale in Fornax (left) 
 and Sculptor (right), resulting from the fit to their velocity dispersion profile under Verlinde's theory. We consider the parameter $a_0$ to be free. As in Test~1, 
 $\beta$ and $\Upsilon_*$ show no correlation between them. However, there is a strong degeneracy between $a_0$ and $\Upsilon_*$. Median values of the posteriors depend on 
 the prior's upper limits, which in this case are given by Eq.~(\ref{eq:prior2}). For reference, the blue line 
represents the theoretical value of the acceleration scale with $a_0=5.4\times 10^{-10}\, {\rm m/s}^{-2}$. This value is consistent at the 2$\sigma$ level for all galaxies, almost independently of the prior's choice.}
 \label{fig:test2}
\end{figure*}

The results are again shown in Table~\ref{tab:results}, where the value of $\Upsilon_*$ is set as the median of the posterior. In this case the mass-to-light ratio is 
unbound from above, and the exact position of the median value strongly depends on the assumed priors for some of the galaxies, in particular for Fornax. Note that 
relaxing the value of $a_0$ allows for smaller stellar mass-to-light ratios, while the values of $\beta$ in both tests are consistent. 
As an example, we show in Figure~\ref{fig:test2} the posterior distributions for Fornax and Sculpture of the three parameters: the acceleration scale, the anisotropy and the mass-to-light ratio.
%(the acceleration scale, the anisotropy, and the stellar mass-to-light ratio) 
We observe that the theoretical value of the acceleration scale used in Test~1, $a_0=5.4\times10^{-10}\,\textrm{m/s}^2$ (blue line reference in 
the figure), is consistent at 2$\sigma$ level for all galaxies. This result holds when a broader prior range in $\Upsilon_*$ is considered. Moreover, all galaxies show a 
strong anticorrelation between the acceleration parameter and the stellar mass-to-light ratio.  This may drive $\Upsilon_*$ to lower values, which are observationally 
preferred, by increasing the value of $a_0$ from the theoretical one. As we mentioned before, this is allowed by Verlinde's model, since we do not know the real deformation 
pattern of the elastic medium by the presence of the dSph galaxies.\footnote{Note that the value preferred by the internal
kinematics of the eight classical dSph's is slightly closer to $a_0=6.5\times10^{-10}\,\textrm{m/s}^2$, consistent with today's Hubble parameter 
(including all matter components). However, we believe that the value in~(\ref{a0planck}), which corresponds to the DE contribution only, is closer to the supporting arguments given by Verlinde about the microstates
that build up dS spacetime.}
Moreover, there is no reason a priori to believe that the strain structure should be universal, 
so each dSph may present a different effective value of $a_0$.

\begin{figure*}
 \includegraphics[width=\textwidth]{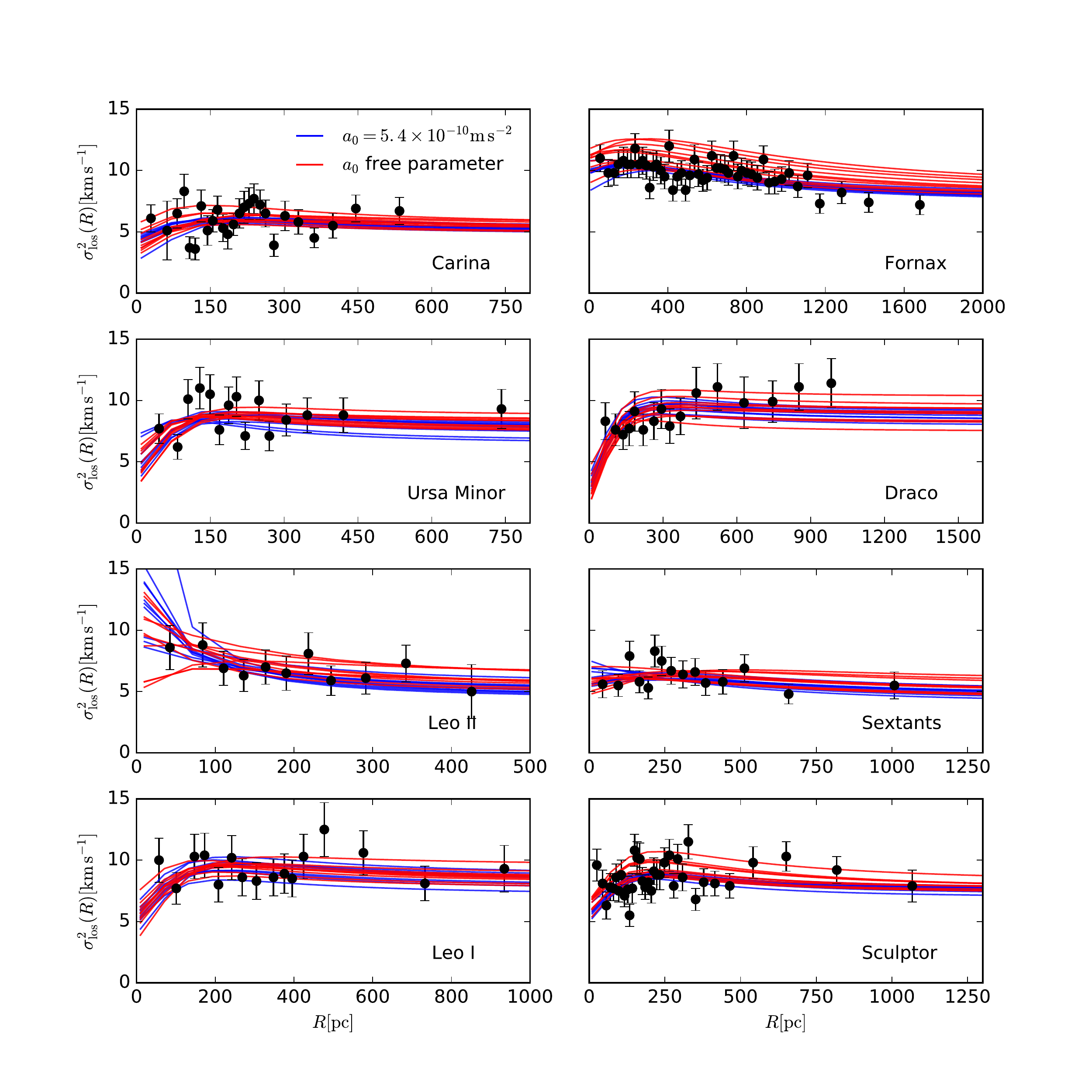}  
 \caption{Empirical velocity dispersion profiles for the eight classical dSph satellites of the Milky Way. Solid lines denote 10 random combinations selected from the 
 resulting posterior distribution in Test~1 (blue lines) and Test~2 (red lines) of Section~\ref{sec.results}, respectively. Both models can fit the data equally well, 
 but from Table~\ref{tab:results}, it is possible to identify that the fits obtained leaving $a_0$ as a free parameter (Test~2) allows for smaller mass-to-light ratios, which are preferred from stellar population and formation history studies.}
 \label{fig:alldisp}
\end{figure*}

In Figure~\ref{fig:alldisp}, we show the empirical velocity dispersion for the eight classical dSph's (black points in the figure) together with 10 random realisations 
selected from the resultant posterior distribution of Test~1 (blue lines), and Test~2 (red lines). Note that in both cases the model describes equally well the data, 
but in Test~2 each line corresponds to a different combination of mass-to-light ratio and acceleration parameter.

Finally, we would like to comment on the argument made at the beginning of this section about the required mass-to-light ratios for Verlinde's and MOND models to fit the 
data. The fact that MOND needs larger values of the mass-to-light ratio was numerically verified by performing the fits described above, for the same Plummer profile, 
but with the effective mass formula determined by MOND's prescription, and with the same fixed value of $a_0$ as in Test~1. One should notice that, in the literature, 
there is another analysis for the dSph's kinematics using MOND by~\cite{Angus:2008vs}, where different assumptions 
were made. In particular, the author considered a King profile with an anisotropic function which was not constant but a linear function in the radial coordinate. 
While using a different density profile may not change drastically our analysis, the non-constant $\beta$ function could hide deviations in the mass-to-light ratio 
with respect to our findings. However, even after these different assumptions, \cite{Angus:2008vs} finds higher mass-to-light ratios than us using Verlinde's model,
with only two cases having $\Upsilon_*<3$. This result is in agreement with our expectations, as we have extensively argued here. Finally, we do not attempt to scrutinize the 
details of the~\cite{Angus:2008vs} study, since we believe a comparison of Verlinde's model to MOND should be carried out under the same assumptions, as we have done here.

\section{Discussion}\label{sec.discussion}

In this paper, we explore the consequences of a proposal where spacetime emerges as a result of the entanglement structure of an underlying microscopic 
theory. %, with an apparent DM component showing up as the consequence of the existence of standard model particles. 
However, apart from some sketches on how it may look like, there is not such a concrete construction of the microscopic theory. Driven by previous arguments such as those 
by~\cite{Sakharov:1967pk} or~\cite{Jacobson:1995ab}, Erik Verlinde has recently 
argued that within such a proposal, an effective description can capture the key ingredients of the relevant microscopic degrees of freedom that describe the DM phenomena~\citep{Verlinde:2016toy}. 
This effective model is simply the theory of linear elasticity, where the DM component appears as a memory effect of the medium due to the presence of standard model particles. 
At this level, one may even forget about the microscopic derivation, and consider the elastic theory as the concrete proposal behind DM. 

For spherically symmetric isolated massive objects, and assuming an elastic-medium response of maximal deformation, Verlinde identifies the expression in 
Eq.~(\ref{eq.MD}) as codifying the correction to the baryonic mass profile $M_B(r)$ for the observed gravitational field, $M(r)=M_B(r)+M_D(r)$. 
Outside a mass, we can simply look at this theory as a particular realization of MOND. We identify that in the exterior limit the deviation of Verlinde's theory with respect 
to the recently proposed phenomenological interpolation function of~\cite{McGaugh:2016leg} is never larger than $30\%$, and marginally accepted by their observations.

However, within the interior part of a massive object, the gravitational field clearly departs from that of MOND. The key difference is that Verlinde's interpolating function 
contains a radial derivative of the mass profile. Therefore, only a very particular mass profile, if any, in Verlinde's theory will reduce to McGaugh's interpolation formula. None of 
the common density models used to describe galaxies in a cosmological framework is expected to be such a special case. In particular, for the case of a regular density profile one finds 
up to a $100\%$ relative difference in the interior of dSph galaxies with respect to the external region, consistent with current astrophysical observations~\citep{Lelli:2017vgz}.
%with respect to the external region in the interior of dSph galaxies, consistent with current astrophysical observations~\citep{Lelli:2017vgz}.

%One should never forget that 
Physics is an experimental discipline, and only observations can decide if a model is correct or not. By following this line of thought, in 
this paper we just took the expression in Eq.~(\ref{eq.MD}) for the spherically symmetric apparent mass distribution for granted, and show that this effective model can 
describe the internal kinematics of the eight classical dSph satellites of the Milky Way without postulating any kind of dark component. In the analysis we found a degeneracy 
between the acceleration parameter, $a_0$, and the mass-to-light ratio, $\Upsilon_*$, which drives $\Upsilon_*$ to values larger than expected as we approach the theoretical 
value of $a_0=c H_0$.

If one considers Eq.~(\ref{eq.MD}) as a phenomenological model to describe DM, and forgets about the elastic medium, the derivation assumptions, or the microstates motivation, 
then one could argue that Verlinde's proposal has a mild tension in the fittings. As shown in our Table~\ref{tab:results}, Test~1, marginally unacceptable 
high values of the mass-to-light ratio show up for some of the objects, such as Draco, Ursa Minor, and Leo~I. However, once we get back to the assumptions of the elastic 
medium, we find that a maximal deformation was considered. By relaxing this condition, 
which is encoded in Eq.~(\ref{generalformula}), the acceleration scale $a_0$ as free parameter can capture the information about the non-maximal deformation of the elastic 
medium, which will translate on a larger value for $a_0$ than the theoretical one, due to the inequality sign of Eq.~(\ref{generalformula}), and in agreement with our findings. One should notice that the preferred value of $a_0$ for some objects is even larger than today's Hubble constant, suggesting that our choice of taking only the DE contribution is not the cause of this tension. Moreover, a value of $a_0$ slightly larger than our theoretical value is also supported by the rotational curves of larger galaxies, 
$a_0=7.2\pm1.56\times 10^{-10}\,\textrm{m/s}^2$~\citep{McGaugh:2016leg,1991MNRAS.249..523B}, although it is important to bear in mind that a non-spherically symmetric fitting formula in Verlinde's proposal 
is still missing. Other astrophysical and modelling systematics could be getting into preferring a higher value of $a_0$ than $cH_0$, but this is beyond our present 
expectations; see \citep{Gonzales-Morales:2016mkl} for discussions on the subject. 
%It is important to say that we have not, in any sense, confronted this hypothesis against the standard DM one.  

In order to improve future analysis of the study presented here, it is important to stress the nature of all the different assumptions made within this work. We considered 
that dSph galaxies can be described as isolated objects with spherical symmetry. Of course these systems are not really isolated, due to their proximity to the Milky Way. 
We have compared the value of the exterior gravitational field due to the Milky Way, $g_\textrm{MW}$, to that of Verlinde's model, $g$, in the different objects. In order to estimate $g_\textrm{MW}$ 
one would need the complete theory, but we can argue that $g_\textrm{MW}\lesssim (170\,\textrm{km/s})^2/R_{\textrm{MW}}$, where the upper bound represents the terminal velocity 
of objects orbiting the Milky Way in the galactic plane at a distance $R_{\textrm{MW}}$ to the centre of the galaxy. In Table~\ref{tab:results} we report the ratio 
$g_\textrm{MW}/g$, with $g\sim (a_0 G\Upsilon_* L_V)^{1/2}/r_{\textrm{half}}$, and find that it is always (apart from Sextans, which does not pose any other problem) 
negligible at the level of the precision obtained by our fittings. On the other hand, these spheroidal systems are not really spherical, but they are the highly DM-dominated 
objects for which such an assumption is best justified, since the radial averaged ellipticity for the eight galaxies is $\epsilon_{avg}=0.05$, with a maximum 
of $\epsilon_{max}\sim 0.3$ [except for Ursa Minor, which has $\epsilon=0.56$, as reported in \cite{2012AJ....144....4M}]. 

To conclude, we have shown how Verlinde's emergent gravity model and MOND predict very different relations in the $g$ versus $g_B$ plane for spherically symmetric configurations, 
which may help discriminate between both proposals. In particular, we have studied in detail the inner configuration of the dSph satellites of the Milky Way. However, more precision 
in both, stellar kinematics and mass-to-light ratios, are required in order to favour one model over the other at the dSph's scales, even though Verlinde's proposal points to mass-to-light 
ratios in those objects that are closer to those expected from stellar population~\citep{Bell:2000jt} and formation history~\citep{Kirby:2013wna} studies. In addition, more theoretical work is needed so 
that one can consider non-spherically symmetric objects. 

\section*{Acknowledgments}
We are grateful to Matthew Walker for providing us with the observational data.  
We are also grateful to Nana Cabo, Oscar Loaiza, Octavio Obreg\'on, Jorge Pe\~narrubia, Miguel Sabido, and Erik Verlinde for useful discussions. AXGM acknowledges support from C\'atedras CONACYT and UCMEXUS-CONACYT collaborative project funding.
This work is supported by CONACyT-Mexico grants 182445, 167335, 179208, 179881, and Fronteras de la Ciencia 281, and also by DAIP-Universidad de Guanajuato grants 
1,046/2016 and 878/2016 .

\bibliography{verlinde_dsph}
\bibliographystyle{mnras}

\appendix
\section{Apparent DM masses in Verlinde's and MOND theories}
The mass distribution of baryons in a dSph can be read from the standard formula $M_B(r)=4\pi\int_0^r \nu(r')r'^2dr'$, once the stellar mass density profile $\nu(r)$ 
is specified. For the case of a Plummer profile~(\ref{eq.3density}) the apparent DM mass in Verlinde's theory, $M_D(r)$ from Eq.~(\ref{eq.MD}), is given by
\begin{equation}\label{Mver}
 M_{D}(x) = \frac{\mathcal{N} x^{5/2}}{(1+x^2)^{3/4}}\sqrt{\frac{4+x^2}{1+x^2}}\,,
\end{equation}
where $x=r/r_{\textrm{half}}$, and
\begin{equation}\label{No}
\mathcal{N}=(L r_{\rm half}^2 \Upsilon_*)^{1/2}\sqrt{\frac{a_0}{6\, G}}\,.
\end{equation}
Notice that the first (second) term in $\mathcal{N}$ is observationally (theoretically) defined.
For comparison we also give the corresponding apparent DM mass in the extreme MOND regime, namely
\begin{equation}\label{Mmond}
M_{D}(x)=\frac{\mathcal{N}_M x^{5/2}}{(1+x^2)^{3/4}}\,,
\end{equation}
where this expression has also been obtained using a Plummer profile.
Here $\mathcal{N}_M$ is equal to the expression in Eq.~(\ref{No}) but replacing 
$a_0\rightarrow 6 a_M$.
From equations (\ref{Mver}) and (\ref{Mmond}), the  difference between Verlinde's theory and MOND apparent DM masses is  $\sqrt{(4+x^2)/(1+x^2)}$, which  gives a factor of 
$2$ at the origin, $x=0$, and of $\sqrt{5/2}$ when evaluated at the half-light radius, $x=1$. Note that in both cases one can get the same value for the apparent DM mass contained at the half-light radius if the stellar mass-to-light 
ratio is $5/2$ times greater in the case of MOND than in Verlinde's model. Therefore, it is natural to understand why Verlinde's emergent gravity proposal points to lower values of 
$\Upsilon_*$ than MOND.

\section{MCMC}\label{appB}

Here we provide some details about the MCMC analysis of Section~\ref{sec.results}, for which we used the EMCEE implementation of~\cite{Foreman-Mackey}. In Test~1, we set the acceleration 
scale to its theoretical value in Eq.~(\ref{a0planck}), $a_0=5.4\times 10^{-10} \textrm{m/s}^{2}$, and leave the mass-to-light ratio, $\Upsilon_*$, and the anisotropy, $\beta$, as free 
parameters. In Test~2 we also promote the acceleration scale to a free parameter. We adopted flat priors in both cases, as given in  Eqs.~\eqref{eq:prior1} and~\eqref{eq:prior2}, respectively. 

We used 40 MCMC walkers in Test~1, and 60 in Test~2,
%In each case we used the maximum of the likelihood to initialize 40 MCMC walkers in Test~1, and 60 in Test~2, 
which correspond to 20 times the number of free 
parameters in each case. Initially, each walker is randomly displaced from the maximum likelihood point by a factor of $10^{-2}$. The chains are set to complete 2,000 MCMC steps. 
We have verified that the walkers quickly start to sample a larger region of the parameter space, and converge to a restricted area. We have analysed the time series 
of the chains to look for convergence. Given the small number of free parameters this is enough to see whether the chains have converged or not. This was the case for all galaxies 
in Test~1. For Test~2, we have noticed that convergence may be affected by the prior range used, in combination with the fact that EMCEE was returning large acceptance 
fractions. To solve this issue we adjusted the EMCEE specifications to ensure an acceptance fraction of approximately $0.4$, and decided to use a narrower prior for 
$\Upsilon_*$ in Test~2 than in Test~1. This was important in the case of Fornax, for which the combination of a large acceptance fraction and prior effects could also lead 
to the appearance of a bimodal distribution, when in reality it is an unbounded distribution (lower limit in the acceleration, upper limit in the mass-to-light ratio).

\end{document}